\documentclass[12pt]{article}
\def\slash#1{\setbox0=\hbox{$#1$}#1\hskip-\wd0\hbox to\wd0{\hss\sl/\/\hss}}
\usepackage{epsf, cite, amsmath, amssymb}
\usepackage{epsfig}
\usepackage[all]{xy}
\makeatletter
\renewcommand\section{\@startsection {section}{1}{\z@}%
                                   {-3.5ex \@plus -1ex \@minus -.2ex}%nn
                                   {2.3ex \@plus.2ex}%
                                   {\normalfont\large\bfseries}}
\renewcommand\subsection{\@startsection{subsection}{2}{\z@}%
                                     {-3.25ex\@plus -1ex \@minus -.2ex}%
                                     {1.5ex \@plus .2ex}%
                                     {\normalfont\bfseries}}
\makeatother

\let\non\nonumber

\newcommand{\bea}{\begin{eqnarray}}
\newcommand{\eea}{\end{eqnarray}}
\newcommand{\be}{\begin{equation}}
\newcommand{\ee}{\end{equation}}

\newcommand{\p}{\partial}

\newcommand{\s}{\sigma}

\newcommand{\C}[1]{$(\ref{#1})$}

\begin{document}

\begin{titlepage}

\begin{center}

%\today
%\hfill

%\hfill

\vskip 2 cm
{\Large \bf The $D^4 {\cal{R}}^4$ term in type IIB string theory on $T^2$ and U--duality}\\
\vskip 1.25 cm { Anirban Basu\footnote{email: abasu@ias.edu}
}\\
{\vskip 0.75cm
Institute for Advanced Study, Princeton, NJ 08540, USA\\
}

\end{center}

\vskip 2 cm

\begin{abstract}
\baselineskip=18pt

We propose a manifestly U--duality invariant modular form for the $D^4 {\cal{R}}^4$ interaction in type
IIB string theory compactified on $T^2$. It receives perturbative
contributions upto two loops, and  
non--perturbative contributions from D--instantons and $(p,q)$ string instantons wrapping $T^2$.
We provide evidence for this modular form by showing that the coefficients at tree 
level and at one loop precisely match those obtained using string perturbation theory.  
Using duality, parts of the perturbative amplitude are also shown to match exactly the results 
obtained from eleven dimensional supergravity compactified on $T^3$ at one loop. 
Decompactifying the theory to nine dimensions, we obtain a U--duality invariant 
modular form, whose coefficients at tree level and at one loop agree with
string perturbation theory.

\end{abstract}

\end{titlepage}

\pagestyle{plain}
\baselineskip=18pt

\section{Introduction}

Understanding duality symmetries of string theory is important in order to analyze the dynamics of the
theory beyond its perturbative regime. In particular, analyzing certain protected operators in toroidal
compactifications of type IIB superstring theory which preserve all the thirty two supersymmetries 
has proven useful in this regard. One such protected operator is the four graviton amplitude in 
the effective action of type IIB string theory, which involves various modular forms of the corresponding 
U--duality groups. These interactions which are of the form $D^{2k} {\cal{R}}^4$ where 
$k$ is a non--negative 
integer, are
expected to satisfy certain non--renormalization properties. 
It has been argued that (at least for low values of $k$)
these interactions receive only a few perturbative contributions, as well as non--perturbative contributions.
The ${\cal{R}}^4$ interaction has been
analyzed in various 
dimensions~\cite{Green:1997tv,Green:1997di,Green:1997as,Kiritsis:1997em,Pioline:1997pu,
Pioline:1998mn,Green:1998by,Obers:1998rn,Obers:1999um,Obers:1999es,Pioline:2001jn} 
(see~\cite{Obers:1998fb,Green:1999qt} for reviews). The $D^{2k} {\cal{R}}^4$
interaction has been analyzed for some higher values of $k$ in~\cite{Green:1999pu,Green:2005ba}, while
the non--renormalization properties have been discussed 
in~\cite{Berkovits:2004px,Berkovits:2006vc,Green:2006gt}.

In this paper, we shall focus on some aspects of the four graviton scattering amplitude
in type IIB superstring theory compactified on $T^2$.
This theory has a conjectured $SL(2, \mathbb{Z} )
\times SL(3, \mathbb{Z})$ U--duality symmetry~\cite{Hull:1994ys,Witten:1995ex}. In fact using dualities,
this U--duality symmetry has a natural geometric interpretation
when one considers M theory compactified on $T^3$. The $SL(3, \mathbb{Z})$ factor is the modular group
of $T^3$, while the Kahler structure modulus $T^M$ on $T^3$ defined by
\be \label{defM} T^M = C_3 + i V_3, \ee
transforms as
\be 
T^M \rightarrow \frac{a T^M + b}{c T^M + d}, \ee
under the $SL(2,\mathbb{Z})$ factor, where $a,b,c,d \in \mathbb{Z}$, and $ad -bc =1$. 
In \C{defM}, $C_3$ is the three form gauge potential of M theory, and
$V_3$ is the volume of $T^3$ in the M theory metric.  

From the eight dimensional point of view,  this U--duality symmetry of type IIB string theory
has a more involved interpretation. The eight dimensional theory has
an $SL(2, \mathbb{Z})_\tau$ S--duality symmetry which is inherited from ten dimensions. It acts on 
the ten dimensional complexified coupling
\be \tau = \tau_1 + i \tau_2 = C_0 + i e^{-\phi} \ee
as
\be \tau \rightarrow \frac{a \tau + b}{c \tau + d}, \ee
and on the combination $B_R + \tau B_N$ 
as
\be B_R + \tau B_N \rightarrow \frac{B_R + \tau B_N}{c\tau + d}, \ee
where $B_N (B_R)$ is the modulus from the NS--NS (R--R) two form on $T^2$.
This theory also has an $SL(2,\mathbb{Z})_T$ T--duality symmetry which 
acts on the Kahler structure modulus of $T^2$ defined by
\be 
T = B_N + i V_2 ,\ee    
as
\be T \rightarrow \frac{a T + b}{c T + d}, \ee
where $V_2$ is the volume of $T^2$ in the 
string frame. It also acts on the complex scalar $\rho$ defined by
\be  \rho = - B_R + i \tau_1 V_2 , \ee
as 
\be \rho \rightarrow \frac{\rho}{c \rho + d}, \ee
while leaving the eight dimensional dilaton invariant. Now the $SL(2,\mathbb{Z})_\tau$
and the $SL(2,\mathbb{Z})_T$ transformations can be intertwined and 
embedded into the $SL(3,\mathbb{Z})$ factor of the U--duality group. Also the $SL(2,\mathbb{Z})$ factor 
of the U--duality group acts on the complex structure modulus $U$ of $T^2$ as
\be U \rightarrow \frac{a U + b}{c U + d}. \ee

The ${\cal{R}}^4$ interaction in type IIB string theory on $T^2$ has been analyzed 
in~\cite{Green:1997as,Kiritsis:1997em} directly in string theory, as well as from the point
of view of eleven dimensional supergravity on $T^3$. In the Einstein frame, where the 
metric is U--duality
invariant, the coefficient of the ${\cal{R}}^4$ interaction is given by a modular form of the 
U--duality group, that is modular invariant under $SL(2, \mathbb{Z}) \times SL(3, \mathbb{Z})$ 
transformations. An expression for this modular form has been 
conjectured in~\cite{Green:1997as,Kiritsis:1997em} 
which we shall mention later. 

In this paper, we consider the $D^4 {\cal{R}}^4$ interaction in type IIB string theory 
compactified on $T^2$. By this, we actually mean the interaction
\be (s^2 + t^2 + u^2) {\cal{R}}^4 \ee 
involving the elastic scattering of two gravitons.
We propose a manifestly U--duality invariant modular form that is the coefficient of this interaction
in the Einstein frame. Explicitly, this modular form is given by
\be 
E_{5/2} (M)^{SL(3,\mathbb{Z})}
-8 E_2 (M^{-1})^{SL(3,\mathbb{Z})} E_{2} (U,\bar{U})^{SL(2,\mathbb{Z})}, \ee
where $E_{5/2} (M)^{SL(3,\mathbb{Z})}$ and $E_2 (M^{-1})^{SL(3,\mathbb{Z})}$ are 
$SL(3,\mathbb{Z})$ invariant modular forms in the
fundamental and the anti--fundamental representations of $SL(3,\mathbb{Z})$ respectively 
as we shall discuss below, 
and $E_{2} (U,\bar{U})^{SL(2,\mathbb{Z})}$ is an $SL(2,\mathbb{Z})$ invariant modular form. 

We first consider some systematics of the four graviton amplitude and briefly review the conjectured
modular form for the ${\cal{R}}^4$ interaction in eight dimensions. In the next section, we argue 
for the modular form for the $D^4 {\cal{R}}^4$ interaction. Our arguments are based on the known
modular form for the $D^4 {\cal{R}}^4$ interaction in ten dimensions, 
U--duality invariance, and the perturbative
equality of the amplitude in type IIA and type IIB string theories. Our proposed modular form satisfies
certain non--renormalization properties: it receives perturbative contributions 
only upto two string loops, as well
as an infinite number of non--perturbative contributions coming from D--instantons 
and $(p,q)$ string instantons
wrapping $T^2$. To provide some evidence for the modular form, we next calculate the four graviton 
amplitude in eight dimensions using string perturbation theory at tree level and at one loop, and show 
that it exactly matches the amplitude given by the modular form. We also consider eleven dimensional
supergravity compactified on $T^3$ at one loop, and obtain parts of the perturbative string theory
amplitude, which are in precise agreement with the coefficients given by the modular form. 
In the next section,
we decompactify the theory to nine dimensions, which has a conjectured $SL(2,\mathbb{Z}) 
\times R^+$ U--duality
symmetry. The modular form we obtain for the $D^4 {\cal{R}}^4$ interaction in nine dimensions manifestly
exhibits this U--duality symmetry. To provide further evidence, we calculate the 
four graviton amplitude in nine
dimensions at tree level and at one loop, and obtain precise agreement with the amplitude given by 
the modular form. We end with some comments about the modular form for the $D^4 {\cal{R}}^4$
interaction for toroidal compactifications to lower dimensions.
   
\section{Some systematics of the higher derivative interactions and the 
${\cal{R}}^4$ interaction}

First let us consider the effective action of type IIB string theory in ten dimensions. In particular,
we consider the perturbative contributions to the protected ${\cal{R}}^4$ and the $D^4 {\cal{R}}^4$ 
interactions along with the Einstein--Hilbert term in the string frame. Here ${\cal{R}}^4$ 
stands for the $t_8 t_8 R^4$ interaction~\cite{Green:1981xx,Green:1981yb,D'Hoker:1988ta}, 
and can be expressed entirely in terms of four powers of the Weyl tensor. Dropping 
various irrelevant numerical factors, these terms are given by
\bea \label{mainact}
S \sim \frac{1}{l_s^8} \int d^{10} x \sqrt{-g} e^{-2\phi} R + \frac{1}{l_s^2} \int d^{10} x 
\sqrt{-g} \Big(2 \zeta(3) e^{-2\phi} + \frac{2\pi^2}{3} + \ldots \Big) {\cal{R}}^4 \non
%\non \\ 
\eea
\bea+ l_s^2
\int d^{10} x \sqrt{-g} \Big( 2 \zeta(5) e^{-2\phi} + \frac{4\pi^4}{135} e^{2\phi}
+ \ldots \Big) D^4 {\cal{R}}^4, \eea 
where $\ldots$ are the various non--perturbative corrections coming from D--instantons. 
Now compactifying on $T^2$ of volume $V_2 l_s^2$ in the string frame
and moving to the eight dimensional Einstein frame, we see that \C{mainact} gives us
\bea \label{actein}
S \sim \frac{1}{l_s^6} \int d^{8} x \sqrt{-\hat{g}_8}  \hat{R} +  \int d^8 x \sqrt{-\hat{g}_8}
V_2 \Big(2 \zeta(3) e^{-2\phi} + \frac{2\pi^2}{3} + \ldots \Big) \hat{\cal{R}}^4 \non \\ + l_s^4
\int d^8 x \sqrt{-\hat{g}_8} V_2^{5/3} e^{-4\phi/3} 
\Big( 2 \zeta(5) e^{-2\phi} + \frac{4\pi^4}{135} e^{2\phi} + \ldots \Big) \hat{D}^4 
\hat{\cal{R}}^4 + \ldots,\eea  
where the hat denotes quantities in the eight dimensional 
Einstein frame. Thus from \C{actein}, we see that the 
$SL(2,\mathbb{Z}) \times SL(3,\mathbb{Z})$ invariant modular form for the $
\hat{\cal{R}}^4$ interaction must contain 
\be V_2 \Big(2 \zeta(3) e^{-2\phi} + \frac{2\pi^2}{3} + \ldots \Big) \ee
among other terms. In fact, an expression for this modular 
form has been conjectured in~\cite{Green:1997as,Kiritsis:1997em}. 
In order to write down the manifestly U--duality invariant modular form, we note that the
part of the supergravity action involving the scalars
can be written in the Einstein frame as (we are following the conventions of~\cite{Kiritsis:1997em})
\be \label{act8}
S \sim \frac{1}{l_s^6} \int d^{8} x \sqrt{-\hat{g}_8} \Big( \hat{R} - \frac{\p_\mu U \hat{\p}^\mu 
\bar{U}}{2 U_2^2} + \frac{1}{4} {\rm Tr} (\p_\mu M \hat{\p}^\mu M^{-1}) + \ldots \Big),\ee 
where $M$ is a symmetric matrix with determinant one given by
\be \label{matrix}
M = \nu^{1/3} \begin{pmatrix} 1/\tau_2 & \tau_1/\tau_2 & {\rm Re}(B)/\tau_2 \\ \tau_1/\tau_2& \vert 
\tau \vert^2/\tau_2 & {\rm Re}(\bar{\tau} B)/\tau_2  \\
{\rm Re}(B)/\tau_2 & {\rm Re}(\bar{\tau} B)/\tau_2 & 1/\nu + \vert B \vert^2/\tau_2  \end{pmatrix}, \ee
where $B = B_R + \tau B_N$, and $\nu = (\tau_2 V_2^2)^{-1}$.
In \C{act8}, the matrices $U$ and $M$ parametrize the coset manifolds $SL(2,\mathbb{R})/SO(2)$ and
$SL(3,\mathbb{R})/SO(3)$ respectively, and so we see that the scalar manifold is reducible and 
is given by $SL(2,\mathbb{R})/SO(2) \times SL(3,\mathbb{R})/SO(3)$. The conjectured U--duality 
group is
generated by the transformations $U \rightarrow (a U +b)/(c U +d)$, and $M 
\rightarrow \Omega_2 M \Omega_2^T$, where $a,b,c,d \in \mathbb{Z}$ with $ad-bc =1$, and 
$\Omega_2 \in SL(3,\mathbb{Z})$. 

The conjectured U--duality invariant modular form for the ${\cal{R}}^4$ interaction is given by 
\be \label{modformR4}
E_{3/2} (M)^{SL(3,\mathbb{Z})}  - 2 \pi {\rm log} (U_2 \vert \eta (U) \vert^4), \ee
where
\be E_{3/2} (M)^{SL(3,\mathbb{Z})}  = \sum'_{m_i} \Big( m_i M_{ij} m_j \Big)^{-3/2} , \ee
where $m_i$ are integers, and the sum excludes $\{ m_1, m_2, m_3 \} = \{ 0,0,0 \}$. Here
$E_s (M)^{SL(3,\mathbb{Z})} $ is the $SL(3,\mathbb{Z})$ invariant Eisenstein series of order
$s$ in the fundamental representation of $SL(3,\mathbb{Z})$, defined by \C{expSL3}. 
Also the other term in \C{modformR4} is the $SL(2,\mathbb{Z})$ invariant Eisenstein series
of order one\footnote{The modular form for the ${\cal{R}}^4$ interaction is actually divergent, and
has to be regularized. For the case of the $D^4 {\cal{R}}^4$ interaction, there are no such divergences.}. 
From \C{modformR4}, it follows that the  ${\cal{R}}^4$ interaction receives 
perturbative contributions only at tree level and at one loop,
and non--perturbative contributions coming from D--instantons and $(p,q)$ string instantons wrapping $T^2$.
This correctly reduces to the $\hat{\cal{R}}^4$ interaction in ten dimensions, which is given by $E_{3/2}
(\tau, \bar\tau)^{SL(2,\mathbb{Z})}$, the $SL(2,\mathbb{Z})$ invariant Eisenstein series of
order $3/2$, where the Eisenstein series of order $s$ is defined by \C{expSL2} (see~\cite{Terras} for 
details).

\section{The modular form for the $D^4 {\cal{R}}^4$ interaction}

We now proceed to construct the modular form for the $D^4 {\cal{R}}^4$ interaction.
From \C{actein}, we see that the U--duality invariant modular form for the $\hat{D}^4 
\hat{\cal{R}}^4$ interaction must contain
\be  V_2^{5/3} e^{-4\phi/3} 
\Big( 2 \zeta(5) e^{-2\phi} + \frac{4\pi^4}{135} e^{2\phi} \Big) + \ldots,  \ee
among other terms. These lead to tree level and two loop contributions when converted to the string frame. 
Our aim is to propose an exact expression for this modular form. 

\subsection{The proposed modular form}

The modular form for the $\hat{D}^4 \hat{\cal{R}}^4$ interaction in ten dimensions is given by $E_{5/2}
(\tau, \bar\tau)^{SL(2,\mathbb{Z})}$, the $SL(2,\mathbb{Z})$ invariant Eisenstein series of
order 5/2~\cite{Green:1999pu}.
From the structure of $E_{5/2} (\tau, \bar\tau)$, it follows that the $\hat{D}^4 \hat{\cal{R}}^4$ interaction
receives perturbative contributions only at tree level and at two loops, and 
an infinite number of non--perturbative contributions from D--instantons.  

 Thus, following the conjecture for the ${\cal{R}}^4$ interaction and given the modular form for the
$D^4 {\cal{R}}^4$ interaction in ten dimensions, it is natural to propose that a part of the full
U--duality invariant modular form for the $\hat{D}^4 \hat{\cal{R}}^4$ interaction is given by the order 5/2
Eisenstein series for $SL(3,\mathbb{Z})$ defined by (see \C{expEs})
\bea \label{expE5/2}
E_{5/2} (M)^{SL(3,\mathbb{Z})} &=& 2 ( \tau_2^2 V_2)^{5/3} \zeta (5) + \frac{4}{3}
(\tau_2^2 V_2)^{- 1/3} E_2 (T,\bar{T})^{SL(2,\mathbb{Z})} \non \\&&
+ \frac{8 \pi^2}{3} \tau_2^{4/3} V_2^{5/3} 
\sum_{m_1 \neq 0 ,m_2 \neq 0} 
\Big\vert \frac{m_1}{m_2} \Big\vert^{2} K_{2} (2\pi \tau_2 \vert m_1 m_2
\vert) e^{2\pi i m_1 m_2 \tau_1} \non
%\non \\  
\eea
\bea
&&+ \frac{2 \pi}{3} \tau_2^{-2/3} V_2^{-4/3} 
\sum_{m_1 \neq 0, m_3 \neq 0, m_2}  
\frac{1 + 2 \pi \vert m_3 (m_2 - m_1 \tau ) \vert V_2 }{\vert m_3 \vert^3} 
\non \\ && 
~~~~~~~~~~~~~~~~~~~~~~~~~~~~~\times e^{-2\pi \vert m_3 (m_2 - m_1 \tau ) \vert V_2
+ 2\pi i m_3 (m_1 B_R + m_2 B_N)},  \eea
where we have used
\be K_{3/2} (x) = \sqrt{\frac{\pi}{2 x}} e^{-x} (1 + x^{-1}). \ee

Now one can read off the various perturbative and non--perturbative contributions to the four
graviton amplitude from \C{expE5/2}. 
While the perturbative contributions are given by the first line of \C{expE5/2}, 
the non--perturbative contributions are from D--instantons which are given by the 
second line of \C{expE5/2}, as well as 
from $(p,q)$ string instantons wrapping $T^2$ with $q \neq 0$\footnote{$(1,0)$ is the fundamental string
in our conventions.} which are 
given by the third line of \C{expE5/2}. 

Thus the perturbative contribution to
$E_{5/2} (M)^{SL(3,\mathbb{Z})}$ is given by
\bea \label{needsym}
E_{5/2} (M)^{SL(3,\mathbb{Z})}_{\rm pert} = 2 ( \tau_2^2 V_2)^{5/3} \zeta (5) + \frac{4}{3}
(\tau_2^2 V_2)^{- 1/3} E_{2} (T,\bar{T})^{SL(2,\mathbb{Z})} . \eea

Now using the fact that we are restricting ourselves to the $t_8 t_8 R^4$ part of the amplitude
which involves only the even--even spin structure, we see that the perturbative contribution must be the
same in type IIA and type IIB string theory. In going from type IIA to type IIB string theory, $\tau_2^2 V_2$
is invariant and $U \leftrightarrow T$, and so we add the relevant two loop 
term to \C{needsym} to get a part
of the perturbative piece of the whole amplitude
\bea \label{pertpiece}
% E^{SL(3,\mathbb{Z})_M \times SL(2,\mathbb{Z})_U}_{\rm pert} = 
2 ( \tau_2^2 V_2)^{5/3} \zeta (5) 
+ \frac{4}{3} (\tau_2^2 V_2)^{- 1/3} \Big( E_{2} (T,\bar{T})^{SL(2,\mathbb{Z})} + 
E_{2}  (U,\bar{U})^{SL(2,\mathbb{Z})} \Big) + \ldots . \eea

Now the first two terms in \C{pertpiece} are obtained from the modular form \C{expE5/2}, 
and so we want to find a modular form which has 
\be \label{twoloop}
\frac{4}{3} (\tau_2^2 V_2)^{- 1/3} E_{2} (U,\bar{U})^{SL(2,\mathbb{Z})} \ee
as the two loop contribution. Now let us consider the modular form
\be E_{-1/2} (M)^{SL(3,\mathbb{Z})} E_{2} (U,\bar{U})^{SL(2,\mathbb{Z})} ,\ee
which has
\bea \label{extmod}
&&E_{-1/2} (M)^{SL(3,\mathbb{Z})}_{\rm pert} E_{2} (U,\bar{U})^{SL(2,\mathbb{Z})} \non
\\ &&=
-\Big[ \frac{1}{6} (\tau_2^2 V_2)^{- 1/3} +\frac{\Gamma (-1)}{2} (\tau_2^2 V_2)^{2/3}
E_{-1} (T,\bar{T})^{SL(2,\mathbb{Z})} \Big] E_{2} (U,\bar{U})^{SL(2,\mathbb{Z})} \eea

where we have used $\zeta (-1) = -1/12$. In \C{extmod}, let us consider the second term which naively
might seem problematic. This is because it contains $\Gamma (-1)$ which is infinite, and also 
because the terms in $E_{-1} (T,\bar{T})^{SL(2,\mathbb{Z})}$ which are not exponentially suppressed 
for large $T_2$ vanish
using \C{expSL2} because $\zeta (-2) =0$. However using the relation \C{imprel}, we see this is not
the case and we get a finite answer for this quantity. In fact we get that
\bea \label{addpert}
&&E_{-1/2} (M)^{SL(3,\mathbb{Z})}_{\rm pert} E_{2} (U,\bar{U})^{SL(2,\mathbb{Z})} \non
\\ &&=
-\Big[ \frac{1}{6} (\tau_2^2 V_2)^{- 1/3} +\frac{1}{2 \pi^3} (\tau_2^2 V_2)^{2/3}
E_{2} (T,\bar{T})^{SL(2,\mathbb{Z})} \Big] E_{2} (U,\bar{U})^{SL(2,\mathbb{Z})}. \eea

Now the first term in \C{addpert} is proportional to \C{twoloop} and yields a two loop contribution,
while the second term contributes at one loop.
Thus it is natural to guess that the modular form which yields \C{twoloop} is given by
\bea -8 E_{-1/2} (M)^{SL(3,\mathbb{Z})} E_{2} (U,\bar{U})^{SL(2,\mathbb{Z})} \non \\
= -8 E_2 (M^{-1})^{SL(3,\mathbb{Z})} E_{2} (U,\bar{U})^{SL(2,\mathbb{Z})},
\eea
where we have used \C{relfundantifund} in going from a modular 
form of ${SL(3,\mathbb{Z})}$ in the fundamental
representation to a modular form in the anti--fundamental representation. 
Thus in the Einstein frame, we get the manifestly U--duality invariant interaction in the
type IIB effective action
\be \label{modform}
l_s^4
\int d^8 x \sqrt{-\hat{g}_8} 
\Big[ E_{5/2} (M)^{SL(3,\mathbb{Z})}
-8 E_2 (M^{-1})^{SL(3,\mathbb{Z})} E_{2} (U,\bar{U})^{SL(2,\mathbb{Z})} \Big] \hat{D}^4 
\hat{\cal{R}}^4 . \ee

Converting to the string frame and considering the
perturbative parts, we see that \C{modform} contributes at tree level, and at one and two loops only. 
More explicitly, the perturbative contributions to the effective action are given in the string frame 
by (upto an overall numerical factor)
\bea  \label{pertexp}
l_s^4 \int d^8 x 
\sqrt{-g_8} \sum_{g=0}^2 (V_2^{-1/2} e^{\phi})^{2g -2} F_g (T,U,\bar{T},\bar{U}) D^4 {\cal{R}}^4 ,\eea
where
\bea \label{checkval}
F_0 (T,U,\bar{T},\bar{U}) &=& 2 \zeta (5), \non \\ F_1 (T,U,\bar{T},\bar{U})
&=& \frac{4}{\pi^3} E_{2} (T,\bar{T})^{SL(2,\mathbb{Z})} 
E_{2} (U,\bar{U})^{SL(2,\mathbb{Z})}, \non \\ F_2 (T,U,\bar{T},\bar{U})
&=& \frac{4}{3} \Big( E_{2} (T,\bar{T})^{SL(2,\mathbb{Z})} +
E_{2} (U,\bar{U})^{SL(2,\mathbb{Z})} \Big).\eea

Thus while going from type IIB to type IIA string theory, which involves 
interchanging $U$ and $T$ while leaving
$e^{-2\phi} V_2$ invariant, we see that \C{pertexp} is invariant, 
and so the perturbative contributions to the
IIA and IIB theories are the same. 

Thus, we propose that the U--duality invariant modular 
form for the $\hat{D}^4 \hat{\cal{R}}^4$ interaction is given by 
\be \label{finform}
E (M,U)^{SL(3,\mathbb{Z}) \times SL(2,\mathbb{Z}) } \equiv E_{5/2} (M)^{SL(3,\mathbb{Z})}
-8 E_2 (M^{-1})^{SL(3,\mathbb{Z})} E_{2} (U,\bar{U})^{SL(2,\mathbb{Z})}, \ee
which satisfies non--renormalization 
properties characteristic of BPS saturated operators. As discussed before, it yields only a finite number of 
perturbative contributions, as well as 
an infinite number of non--perturbative contributions involving D--instantons
and $(p,q)$ string instantons, as well as perturbative fluctuations about their backgrounds. It involves
modular forms of $SL(2,\mathbb{Z})_U$ and $SL(3,\mathbb{Z})_M$ which satisfy the Laplace equations
\be \Delta_{SL(2,\mathbb{Z})} E_s (U,\bar{U})^{SL(2,\mathbb{Z})} = 4 U_2^2 \frac{\p^2}{\p U \p \bar{U}} 
E_s (U,\bar{U})^{SL(2,\mathbb{Z})} = s (s -1) E_s (U,\bar{U})^{SL(2,\mathbb{Z})},\ee
and~\cite{Kiritsis:1997em}   
\bea \Delta_{SL(3,\mathbb{Z})} E_s (M)^{SL(3,\mathbb{Z})} &=& 
\Big[ 4 \tau_2^2 \frac{\p^2}{\p \tau \p \bar\tau} 
+ \frac{1}{\nu \tau_2} \vert \p_{B_N} - \tau \p_{B_R} \vert^2 + 3 \p_\nu (\nu^2 \p_\nu)
\Big]E_s (M)^{SL(3,\mathbb{Z})} \non \\ &&= \frac{2s (2s -3)}{3} E_s (M)^{SL(3,\mathbb{Z})}, \eea
on the fundamental domains of $SL(2,\mathbb{Z})_U$ and $SL(3,\mathbb{Z})_M$ respectively. 

\subsection{Evidence using string perturbation theory}

We now provide some evidence for the modular form \C{finform} using superstring perturbation theory. 
We shall show that the
four graviton amplitude in eight dimensions 
at tree level and at one loop precisely gives the values predicted by the modular form. 

The sum of the contributions to the four graviton amplitude at tree level~\cite{Green:1981xx,D'Hoker:1988ta} 
and at one loop~\cite{Green:1981ya,D'Hoker:1988ta} in type II 
string theory compactified on an $n$ dimensional torus $T^n$ is proportional to (we choose
the overall numerical factor to match the coefficients in \C{checkval})\footnote{The calculation actually
yields ${\cal{R}}^4$ at the linearized level.}
\be \label{totalcont}
32 \Big[- V_n e^{-2 \phi} \frac{\Gamma (-l_s^2 s/4) \Gamma (-l_s^2 t/4) 
\Gamma (-l_s^2 u/4)}{ \Gamma (1 + l_s^2 s/4) \Gamma (1 + l_s^2 t/4)
\Gamma (1 + l_s^2 u/4)} + 2\pi I \Big] {\cal{R}}^4,\ee
where $V_n$ is the volume of $T^n$ in the string frame, and 
$I$ is obtained from the one loop amplitude, and is given by
\be \label{defd1}
I = \int_{\cal{F}} \frac{d^2 \Omega}{\Omega_2^2} Z_{lat} F(\Omega,\bar\Omega) ,\ee
where $\cal{F}$ is the fundamental domain of $SL(2,\mathbb{Z})$, and $d^2 \Omega = d\Omega d \bar\Omega/2$. 
The relative coefficient between the tree level and the one loop
terms in \C{totalcont} is fixed using unitarity~\cite{Sakai:1986bi}. 
In \C{defd1}, the lattice factor $Z_{lat}$ which depends on the moduli is given by
\be Z_{lat} = V_n \sum_{m_i,n_i \in \mathbb{Z}} e^{-\frac{\pi}{\Omega_2} \sum_{i,j} (G + B_N)_{ij}
(m_i + n_i \Omega) (m_j + n_j \bar\Omega)} ,\ee
where $i,j = 1,\dots,n$. Specializing to the case of $T^2$, we get that~\cite{Dixon:1990pc}
\bea \label{deflattice}
Z_{lat} &=& V_2 \sum_{m_1,m_2,n_1,n_2 \in \mathbb{Z}} e^{-\frac{\pi}{\Omega_2} \sum_{i,j} (G + B_N)_{ij}
(m_i + n_i \Omega) (m_j + n_j \bar\Omega)} \non \\ &=& V_2 \sum_{A \in {Mat} (2 \times 2, \mathbb{Z})}
{\rm exp} \Big[ -2\pi i T ({\rm det} A) - \frac{\pi T_2}{\Omega_2 U_2} 
\Big\vert \begin{pmatrix} 1 & U \end{pmatrix}
A \begin{pmatrix} \Omega \\ 1 
\end{pmatrix} \Big\vert^2 \Big], \eea
where 
\be G_{ij} = \frac{T_2}{U_2} \begin{pmatrix} 1 & U_1 \\ U_1 & \vert U \vert^2 
\end{pmatrix}. \ee

%and

%\be 
%A =  \begin{pmatrix} m_1 & n_1  \\ m_2 & n_2 
%\end{pmatrix}. \ee

Also the dynamical factor $F(\Omega,\bar\Omega)$ in \C{defd1} is given by
\be  \label{factF}
F(\Omega,\bar\Omega) = \int_{\cal{T}} \prod_{i=1}^3 \frac{d^2 \nu_i}{\Omega_2} 
(\chi_{12} \chi_{34})^{l_s^2 s} (\chi_{14} \chi_{23})^{l_s^2 t} (\chi_{13} \chi_{24})^{l_s^2 u}. \ee

In \C{factF}, $\nu_i$ ($i=1,\ldots,4$) are the positions of insertions of
the four vertex operators on the toroidal 
worldsheet, 
and $\nu_4$ has been set equal to $\Omega$ using conformal invariance. Also $d^2 \nu_i = d \nu_i^R d\nu_i^I$,
where $\nu_i^R$ ($\nu_i^I$) are the real (imaginary) parts of $\nu_i$. The integral over $\cal{T}$ is over
the domain ${\cal{T}} = \{ -1/2 \leq \nu_i^R < 1/2 ,  0 \leq \nu_i^I < \Omega_2 \}$.
Finally, ${\rm ln } \chi_{ij} (\nu_i - \nu_j ; \Omega)$ is the scalar Green function between the points
$\nu_i$ and $\nu_j$ on the toroidal worldsheet and is given by
\be \label{proptorus}
{\rm ln } \chi (\nu ; \Omega) = \frac{1}{4\pi} \sum_{(m,n) \neq (0,0)} \frac{\Omega_2}{\vert m \Omega 
+ n \vert^2} e^{\pi [ \bar\nu (m\Omega + n )-\nu(m\bar\Omega + n)]/\tau_2} + \frac{1}{2} {\rm ln} \Big\vert 
(2\pi)^{1/2}\eta(\Omega) \Big\vert^2.\ee

In \C{proptorus}, the last term which is the zero mode does not contribute to the
on--shell amplitude and hence can be dropped. In evaluating \C{factF} to fourth order in the momenta,
we use the relation~\cite{Green:1999pv}
\bea \int_{\cal{T}} \frac{d^2 \nu_i d^2 \nu_j}{\Omega_2^2}  [ {\rm ln } \chi (\nu_i - \nu_j ; 
\Omega)]^2 = \frac{1}{16 \pi^2} \sum_{(m,n) \neq (0,0)} \frac{\Omega_2^2}{\vert m \Omega + n \vert^4}
= \frac{1}{16 \pi^2} E_{2}
(\Omega, \bar\Omega)^{SL(2,\mathbb{Z})} ,\eea
which can be deduced using \C{proptorus} with the zero mode term removed.
Thus, expanding to fourth order in the momenta, the total contribution of the
tree level term and the one loop term in \C{totalcont} gives
\be \label{needcont}
\Big[ 2 \zeta (5) V_2 e^{-2 \phi} +\frac{4}{\pi} \int_{{\cal{F}}_L} \frac{d^2 \Omega}{\Omega_2^2} 
Z_{lat} E_{2}
(\Omega, \bar\Omega)^{SL(2,\mathbb{Z})} \Big] l_s^4 (s^2 + t^2 + u^2 ) {\cal{R}}^4 .\ee

In \C{needcont}, note that the one loop contribution has been integrated 
over the restricted fundamental domain 
${\cal{F}}_L$ of $SL(2,\mathbb{Z})$, which is obtained from $\cal{F}$ by restricting to $\Omega_2 
\leq L$. This is necessary to separate the analytic parts of the amplitude from the non--analytic parts
(see~\cite{Green:1999pv} for a detailed discussion).
The integral over ${\cal{F}}_L$ gives both finite and divergent terms to the amplitude in the 
limit $L \rightarrow \infty$. The terms which are finite in this limit are the 
analytic parts of the amplitude. 
The parts which diverge in this limit cancel in the whole amplitude when the contribution from the part of 
the moduli space $\cal{F}$ with $\Omega_2 > L$ is also included. In addition to 
these divergences which cancel,
the contribution from $\cal{F}$ with $\Omega_2 > L$ also gives the various 
non--analytic terms in the amplitude.
Keeping this in mind, we shall consider only the contributions which are finite in the limit 
$L \rightarrow \infty$ in \C{needcont}, and drop all divergent terms. In the calculations below,
we shall see that the domain of integration ${\cal{F}}$ shall often be changed to the upper half plane or a 
strip. Then truncating to ${\cal{F}}_L$ to calculate the analytic terms cannot be done when the
integration over ${\cal{F}}_L$ produces divergences of the form ${\rm ln} L$~\cite{Green:1999pv}. However,
from \C{needcont}, using the expression for $E_{2} (\tau, \bar\tau)^{SL(2,\mathbb{Z})}$, we see that there
are no logarithmic divergences, and so this is not a problem for us. 

We write
\be \label{totint}
\int_{{\cal{F}}_L} \frac{d^2 \Omega}{\Omega_2^2} 
Z_{lat} E_{2}
(\Omega, \bar\Omega)^{SL(2,\mathbb{Z})} = I_1 + I_2 + I_3 ,\ee
where $I_1, I_2$, and $I_3$ are the contributions from the zero orbit, the non--degenerate orbits and 
the degenerate orbits of $SL(2,\mathbb{Z})$ respectively~\cite{Dixon:1990pc} 
(also see~\cite{Bachas:1997mc}). 
Now, from \C{expSL2}, we get that
\bea \label{needdef}
E_{2} (\Omega,\bar{\Omega})^{SL(2,\mathbb{Z})} 
&=& 2 \zeta(4) \Omega_2^2 +  \frac{\pi \zeta (3)}{\Omega_2} 
\non \\ &&+ 2 \pi^2 \sqrt{\Omega_2} \sum_{m_1 \neq 0 ,m_2 \neq 0} 
\Big\vert \frac{m_1}{m_2} \Big\vert^{3/2} K_{3/2} (2\pi \Omega_2 \vert m_1 m_2
\vert) e^{2\pi i m_1 m_2 \Omega_1}. \eea
 
We now calculate the contributions to \C{totint} from the various orbits. In doing the integrals, we 
frequently make use of the definition
\be K_s (x) = \frac{1}{2} \Big(  \frac{x}{2} \Big)^s \int_0^\infty \frac{dt}{t^{s+1}} e^{-t - x^2/4t} . \ee 

{\bf{(i)}} The contribution from the zero orbit involves setting $A =0$ in \C{deflattice} leading to
\be \label{I1cont}
I_1 = V_2 \int_{{\cal{F}}_L} \frac{d^2 \Omega}{\Omega_2^2} E_{2}
(\Omega, \bar\Omega)^{SL(2,\mathbb{Z})}  =0 , \ee
upto $L$ dependent terms. In doing this integral, we use the fact that 
\be  \Delta_{SL(2,\mathbb{Z})} E_{2}
(\Omega, \bar\Omega)^{SL(2,\mathbb{Z})} = 2 E_{2}
(\Omega, \bar\Omega)^{SL(2,\mathbb{Z})}. 
\ee

Thus \C{I1cont} 
picks up contributions only from the boundary of ${\cal{F}}_L$ which is at
$\Omega_2 = L$. We do not get any term which is finite as $L \rightarrow 
\infty$, and thus \C{I1cont} vanishes~\cite{Green:1999pv}. In fact, this is the reason why the one 
loop contribution to the 
${D}^4 {\cal{R}}^4$ interaction vanishes in ten dimensions. 

The contributions from the non--degenerate and degenerate orbits yield finite pieces when $L \rightarrow 
\infty$, and so we directly integrate over ${\cal{F}}$ rather than ${\cal{F}}_L$ in the expressions below.

{\bf{(ii)}} The contribution from the non--degenerate orbits involves setting
\be 
A =  \begin{pmatrix} k & j  \\ 0 & p 
\end{pmatrix} \ee
in \C{deflattice}, where $ k > j \geq 0, p \neq 0$, and changing the domain of integration to be the 
double cover of the upper half plane. This leads to
\bea \label{I2cont}
I_2 &=& 2 V_2 \int_{-\infty}^\infty d \Omega_1 \int_0^\infty \frac{d\Omega_2}{\Omega_2^2} E_{2}
(\Omega, \bar\Omega)^{SL(2,\mathbb{Z})}
\sum_{k > j \geq 0, p \neq 0} e^{-2\pi i T kp - \frac{\pi T_2}{\Omega_2 U_2} 
\vert k\Omega + j + pU \vert^2} \non \\ &=& 2 \Big( 2 \zeta (4) U_2^2 
+ \frac{\pi \zeta (3)}{U_2} \Big) \sqrt{T_2} \sum_{p \neq 0 ,k \neq 0} 
\Big\vert \frac{p}{k} \Big\vert^{3/2} K_{3/2} (2\pi T_2 \vert p k
\vert) e^{2\pi i p k T_1} \non \\ && + 4\pi^2 \sqrt{U_2 T_2} 
{\hat\sum}_{m \neq 0, n \neq 0, p \neq 0, q \neq 0}
\Big\vert \frac{m}{n} \Big\vert^{3/2} K_{3/2} (2\pi  \vert pq
\vert U_2) K_{3/2} (2\pi  \Big\vert \frac{mnp}{q} \Big\vert T_2) \non \\
&&~~~~~~~~~~~~~~~~~~~~~~~~~~~~~~~~~~~~~~~~~~~~~~~~~~~~~~~~~\times e^{2\pi i p (q U_1 + mnT_1/q)}.
\eea

The three terms in \C{I2cont} are obtained from the three terms in \C{needdef} in the order the expressions
are written, and we have also used 
\be K_{1/2} (x) = \sqrt{\frac{\pi}{2 x}} e^{-x} . \ee

In evaluating the integrals in \C{I2cont}, we always first do the $\tau_1$ integral, then sum over $j$, and 
then finally do the $\tau_2$ integral. In \C{I2cont}, the last term involves a restricted sum
which involves integers $m,n$, and $q$ such that $(mn)/q$ is an integer. Now defining
\be pq = \hat{p} \hat{q} , \quad \frac{mnp}{q} = \hat{m} \hat{n} , \quad \frac{m}{n} = \frac{\hat{p} 
\hat{m}}{\hat{q} \hat{n}} , \ee
where $\hat{m} , \hat{n}, \hat{p}$, and $\hat{q}$ are non--zero integers, we get an unrestricted sum
\bea {\hat\sum}_{m \neq 0, n \neq 0, p \neq 0, q \neq 0}
\Big\vert \frac{m}{n} \Big\vert^{3/2} K_{3/2} (2\pi  \vert pq
\vert U_2) K_{3/2} (2\pi  \Big\vert \frac{mnp}{q} \Big\vert T_2) 
 e^{2\pi i p (q U_1 + mnT_1/q)} ~~~~~~\non \\ = \Big\{ \sum_{\hat{p} \neq 0, \hat{q} \neq 0} 
\Big\vert \frac{\hat{p}}{\hat{q}} \Big\vert^{3/2} K_{3/2} (2\pi  \vert \hat{p} \hat{q}
\vert U_2)  e^{2\pi i \hat{p} \hat{q} U_1} \Big\} \Big\{
\sum_{\hat{m} \neq 0, \hat{n} \neq 0} \Big\vert \frac{\hat{m}}{\hat{n}} \Big\vert^{3/2}
K_{3/2} (2\pi  \vert \hat{m} \hat{n} \vert T_2) e^{2\pi i \hat{m} \hat{n} T_1}  \Big\}. \eea
 
Thus we get that
\be \label{I2ConT}
I_2 = 2 \sqrt{T_2} E_{2} (U, \bar{U})^{SL(2,\mathbb{Z})} \sum_{p \neq 0 ,k \neq 0} 
\Big\vert \frac{p}{k} \Big\vert^{3/2} K_{3/2} (2\pi T_2 \vert p k
\vert) e^{2\pi i p k T_1}.\ee

{\bf{(iii)}} The contribution from the degenerate orbits involves setting
\be 
A =  \begin{pmatrix} 0 & j  \\ 0 & p 
\end{pmatrix} \ee
in \C{deflattice} such that $(j,p) \neq (0,0)$, and changing the domain of integration to be the strip 
$\Omega_2 > 0, \vert \Omega_1 \vert < 1/2$, leading to
\bea \label{degenorb}
I_3 &=& V_2 \int_{-1/2}^{1/2} d \Omega_1 \int_0^\infty \frac{d\Omega_2}{\Omega_2^2} E_{2}
(\Omega, \bar\Omega)^{SL(2,\mathbb{Z})} \sum_{(j,p) \neq (0,0)} 
e^{-\frac{\pi T_2}{\Omega_2 U_2} \vert j + pU \vert^2}\non \\
&=& \frac{1}{\pi^2} \Big( 2 \zeta (4) T_2^2
+ \frac{\pi \zeta (3) }{T_2} \Big)  E_{2} (U, \bar{U})^{SL(2,\mathbb{Z})} , \eea
where have used \C{relSL2} with $s = -1$, for $E_s(U, \bar{U})^{SL(2,\mathbb{Z})}$. Note that
the contribution of the last term in \C{needdef} to \C{degenorb} vanishes because of the $\Omega_1$ integral.

Thus from \C{I1cont}, \C{I2ConT}, and \C{degenorb}, we get that 
\be \int_{{\cal{F}}_L} \frac{d^2 \Omega}{\Omega_2^2} 
Z_{lat} E_{2}
(\Omega, \bar\Omega)^{SL(2,\mathbb{Z})} = \frac{1}{\pi^2} E_{2} (U, \bar{U})^{SL(2,\mathbb{Z})} 
E_{2} (T, \bar{T})^{SL(2,\mathbb{Z})} , \ee
which when substituted in \C{needcont} precisely gives the coefficients $F_0$
and $F_1$ in \C{checkval}. Thus we get a non--trivial consistency check of the proposed modular form
using string perturbation theory. 
  
We will have nothing to say about the two loop calculation of the amplitude apart from making a minor 
comment. Dropping numerical factors, the relevant term in the two loop amplitude 
involving four powers of momenta is given by~\cite{D'Hoker:2005ht}
\be \label{genus2}
e^{2\phi} (s^2 + t^2 + u^2)
{\cal{R}}^4 \int_{{\cal{M}}_2} \frac{ \vert d^3 \Omega \vert^2}{({\rm det Im} \Omega)^3} Z_{lat},  \ee
where $\Omega_{AB}$ is the period matrix, $Z_{lat}$ is the lattice factor given by
\be \label{genus2deflat}
Z_{lat} = V_2 \sum_{m_{iA}, n_j^B \in \mathbb{Z}} e^{-\pi \sum_{i,j} (G + B_N)_{ij}
(m_{iA} + \Omega_{AB} n_i^B ) ({\rm Im} \Omega)^{AC} (m_{jC} + \bar\Omega_{CD} n_j^D )},\ee
and the integral is over ${\cal{M}}_2$, the 
fundamental domain of $Sp(4,\mathbb{Z})$ (see~\cite{Klingen}, for example). Unlike the one loop calculation,
probably one does not need to restrict the integral \C{genus2} to a restricted fundamental domain of
$Sp(4,\mathbb{Z})$. This is because the problematic term involves configurations where the
fundamental string worldsheet has vanishing winding which gives $Z_{lat} =V_2$ in \C{genus2deflat}. 
However, 
this term gives the volume of the fundamental domain of $Sp(4,\mathbb{Z})$ and is 
finite~\cite{Siegel}.
This is the reason why the two loop contribution to $D^4 {\cal{R}}^4$ is non--vanishing in ten 
dimensions. The other contributions which involve non--trivial 
$Z_{lat}$ are expected to converge giving a finite answer.
It would be interesting to find the two loop coefficient and see if it agrees with $F_2$ in \C{checkval}. 

\subsection{Evidence using eleven dimensional supergravity on $T^3$ at one loop}

We now provide some evidence for \C{checkval} using the four graviton amplitude in eleven dimensional 
supergravity compactified on $T^3$. It is known that the $D^4 {\cal{R}}^4$ interaction 
receives contributions only from one and two loops\footnote{The three 
loop contribution has leading dependence
$D^6 {\cal{R}}^4$~\cite{Bern:2007hh}.}. We shall consider only the one loop amplitude
and show that it reproduces some terms in \C{checkval}\footnote{In this section, loops refer to
spacetime loops in eleven dimensional supergravity on $T^3$. We shall refer to the worldsheet expansion 
of string perturbation theory as the genus expansion.}. 

The one loop four graviton amplitude is given 
by~\cite{Green:1982sw,Russo:1997mk,Green:1997ud,Green:1999pu}     
\be \label{d11}
A_4 = \frac{\kappa_{11}^4}{(2 \pi)^{11}} \hat{K} [I(S,T) + I(S,U) + I (U,T)] , \ee
where $\hat{K}$ involves the ${\cal{R}}^4$ interaction at the linearized level, and 
\be \label{imprel2}
I(S,T) = \frac{2 \pi^4}{l_{11}^3 {V_3}} \int_0^\infty \frac{d \s}{\s} \int_0^1 d 
\omega_3 \int_0^{\omega_3} d \omega_2 \int_0^{\omega_2} d \omega_1 \sum_{\{ l_1,l_2,l_3 \}} 
e^{- G^{IJ} l_I l_J \s/l_{11}^2 -  Q(S,T;\omega_r) \s},\ee 
where $Q(S,T;\omega_r) = -S \omega_1 (\omega_3 - \omega_2) - T (\omega_2 - \omega_1) (1 
- \omega_3)$~\footnote{Note that $\s$ has dimensions of $({\rm length})^2$.}. 
Denoting the torus directions as $1,2$, and $3$, we choose $G_{11} = R_{11}^2$ to be the metric along the 
M theory circle, thus $R_{11} = e^{2\phi^A/3}$.
Though we need the $(s^2 + u^2 + t^2) {\cal{R}}^4$ term, we shall later find it useful to 
extract a part of the momentum independent amplitude from \C{d11} in order
to fix normalizations. This is given by
\bea \label{zeromom}
A_4 (S= T =U =0) &=& \frac{\kappa_{11}^4 \hat{K}}{(2 \pi)^{11}} \cdot 
\frac{\pi^4}{l_{11}^3 {V}_3} \int_0^\infty \frac{d \s}{\s}
\sum_{\{ l_1,l_2,l_3 \}} e^{- G^{IJ} l_I l_J \s/l_{11}^2 } \non \\ &=& \frac{\kappa_{11}^4 
\hat{K}}{(2 \pi)^{11}} \cdot \pi^4 \int_0^\infty \frac{d \s}{\s^{5/2}}
\sum_{\{ \hat{l}_1, \hat{l}_2 , \hat{l}_3 \}} e^{- \frac{\pi G_{IJ} \hat{l}_I \hat{l}_J 
l_{11}^2}{ \s} } ,  \eea
where we have done Poisson resummation using \C{needresum}. Considering the $\hat{l}_1 \neq 0,
\hat{l}_2 = \hat{l}_3 =0$ piece, \C{zeromom} gives~\cite{Green:1997as}
\be \label{partcont1}
A_4 (S= T =U =0) =  \frac{\kappa_{11}^4 \hat{K}}{(2 \pi)^{11} l_{11}^3} \Big[ \pi^3 \zeta (3)
e^{-2\phi^A}+ \ldots \Big] .\ee

As an aside, 
note that the contribution of the non--analytic part of the amplitude involves setting $l_I =0$ in
\C{imprel2} leading to (see~\cite{Green:1997ud} for relevant discussion)
\bea I(S,T)_{\rm non-anal} &=& \frac{2 \pi^4}{l_{11}^3 {V_3}} \int_0^\infty \frac{d \s}{\s} \int_0^1 d 
\omega_3 \int_0^{\omega_3} d \omega_2 \int_0^{\omega_2} d \omega_1 
(e^{-  Q(S,T;\omega_r) \s } -1) \non \\ &=& -\frac{2 \pi^4}{l_{11}^3 {V_3}} \int_0^1 d 
\omega_3 \int_0^{\omega_3} d \omega_2 \int_0^{\omega_2} d \omega_1 {\rm ln } (-Q(S,T;\omega_r)). \eea

We now consider the analytic part
of \C{d11} which involves
\bea I (S,T)_{\rm anal} &=& \frac{2\pi^4}{l_{11}^3 V_3} \sum_{n=2}^\infty  \frac{{\cal{G}}_{ST}^n}{n !} 
\sum_{( l_1,l_2,l_3 ) \neq ( 0,0,0 )} \int_0^\infty \frac{d\s}{\s^{1-n}} 
e^{- G^{IJ} l_I l_J \s/l_{11}^2} \non \\ &=& 
\frac{2\pi^4 l_{11}^{2n -3}}{V_3} \sum_{n=2}^\infty  \frac{{\cal{G}}_{ST}^n}{n} 
E_n (G^{-1})^{SL(3,\mathbb{Z})} ,\eea
where 
\be {\cal{G}}_{ST}^n = \int_0^1 d 
\omega_3 \int_0^{\omega_3} d \omega_2 \int_0^{\omega_2} d \omega_1 
\Big( -Q(S,T;\omega_r) \Big)^n ,\ee
and we have used \C{expEanti}. Focussing on the $n=2$ contribution, we see that
\be \label{poisson8d}
I (S,T)_{\rm anal}^{n=2} = \pi^6 {\cal{G}}_{ST}^2 \int_0^\infty  d\s \s^{-1/2} 
\sum_{( \hat{l}_1, \hat{l}_2, \hat{l}_3 ) \neq 
( 0,0,0 )}  e^{- \pi G_{IJ} \hat{l}_I \hat{l}_J l_{11}^2/\s} .\ee

We shall be interested only in those terms in \C{poisson8d} thats lead to the perturbative string
contributions given in \C{checkval}. To evaluate \C{poisson8d}, we split the sum over $\hat{l}_I$
into two parts: (i) $(\hat{l}_2, \hat{l}_3) = (0,0), \hat{l}_1 \neq 0$, and (ii) $(\hat{l}_2, \hat{l}_3) 
\neq (0,0)$, $\hat{l}_1$ arbitrary, and call these contributions $I (S,T)_{\rm anal}^1$ and 
$I (S,T)_{\rm anal}^2$ respectively. We get that
\be \label{partcont2} I (S,T)_{\rm anal}^1 = \frac{\pi^7}{3} {\cal{G}}_{ST}^2 l_{11} e^{2\phi^A/3} . \ee 

To calculate $I (S,T)_{\rm anal}^2$, we Poisson resum on $\hat{l}_1$ to go back to $l_1$, to get 
\bea \label{compexpr}
I (S,T)_{\rm anal}^2 = \frac{\pi^6 {\cal{G}}_{ST}^2}{l_{11} R_{11}} 
\sum_{(\hat{l}_2, \hat{l}_3) \neq (0,0), l_1} \int_0^\infty d \s {\rm exp} \Big[
\frac{2\pi i l_1}{G_{11}}  \Big( G_{12} \hat{l}_2 + G_{13}
\hat{l}_3 \Big) - \frac{\pi l_1^2 \s}{l_{11}^2 R_{11}^2} \non
\\ - \frac{\pi l_{11}^2}{\s} \Big\{ 
\hat{l}_2^2 \Big( G_{22} - \frac{G_{12}^2}{G_{11}} \Big) 
+ \hat{l}_3^2 \Big( G_{33} - \frac{G_{13}^2}{G_{11}} 
\Big) + 2 \hat{l}_2 \hat{l}_3 \Big( G_{23} - \frac{G_{12} G_{13}}{G_{11}} \Big)  \Big\} \Big].
\eea

We next split \C{compexpr} into two parts: $(\hat{l}_2, \hat{l}_3) \neq (0,0) , l_1 = 0$ which we
call $I (S,T)_{\rm anal}^{2,0}$, and $(\hat{l}_2, \hat{l}_3) \neq (0,0) , l_1 \neq 0$, which we 
call $I (S,T)_{\rm anal}^{2,1}$. 
To express them in terms of quantities in type IIA string theory,
we use the IIA string frame metric
\be g_{i-1,j-1}^A = R_{11} \Big( G_{ij} - \frac{G_{1i} G_{1j}}{G_{11}} \Big),\ee
where $i,j = 2,3$. Also, the moduli from the R--R one form potentials along $T^2$ in type IIA are given by
\be A_{i-1} = \frac{G_{1i}}{G_{11}} ,\ee 
where $i,j = 2,3$. Finally, the complex structure $U$ of $T^2$ of volume $T_2$ is given by
\be U = \frac{1}{g_{22}^A} (g_{23}^A + i\sqrt{{\rm det g^A}}).\ee

Thus we get that
\be \label{partcont3} I (S,T)_{\rm anal}^{2,0} = \frac{\pi^4 l_{11}}{R_{11}^2} T_2 {\cal{G}}_{ST}^2 
E_2 (U,\bar{U})^{SL(2,\mathbb{Z})} . \ee

Note that $I (S,T)_{\rm anal}^{2,1}$ gives non--perturbative contributions which are not
relevant for \C{checkval}, and so we shall neglect them\footnote{This
contribution is given by
\be I (S,T)_{\rm anal}^{2,1} = 2 \pi^6 {\cal{G}}_{ST}^2 \frac{l_{11}}{\sqrt{R_{11}}} \sqrt{\frac{T_2}{U_2}}
\sum_{(\hat{l}_2, \hat{l}_3) \neq (0,0) , l_1 \neq 0} \frac{\vert \hat{l}_2 +
\hat{l}_3 U \vert}{\vert l_1 \vert} K_1 \Big( 2\pi e^{-\phi^A}\sqrt{\frac{T_2}{U_2}} 
\vert \hat{l}_2 + \hat{l}_3 U \vert \vert l_1 \vert \Big) e^{2\pi i l_1 \hat{l}_i A_i}.\ee}.

Thus from \C{partcont1}, \C{partcont2}, and  \C{partcont3}, we get that
\be \label{needcompl}
A_4  =  \frac{\kappa_{11}^4 \hat{K}}{(2 \pi)^{11} l_{11}^3} \Big[ \pi^3 \zeta (3)
e^{-2\phi^A} + \Big\{ \frac{\pi^4}{6!} T_2 E_2 (U,\bar{U})^{SL(2,\mathbb{Z})} 
+ \frac{\pi^7}{3 \cdot 6!} e^{2\phi^A} \Big\} l_s^4 (s^2 + t^2 + u^2)+ \ldots \Big] ,\ee
where we have used $l_{11} = e^{\phi^A/3} l_s$, and 
\be {\cal{G}}_{ST}^2 + {\cal{G}}_{SU}^2 + {\cal{G}}_{UT}^2 = \frac{1}{6!} (s^2 +  t^2 + u^2) .\ee

From \C{needcompl}, we see that the one loop supergravity amplitude contributes only at genus one and genus
two in the $D^4 {\cal{R}}^4$ interaction in type IIA string theory. 
However, given the genus zero ${\cal{R}}^4$ interaction in
\C{needcompl}, we can fix the normalization of the genus zero $D^4 {\cal{R}}^4$ interaction using
\C{totalcont}. The genus zero interaction in \C{totalcont} is proportional to
\be T_2 e^{-2 \phi^A} \Big( \zeta (3) + \frac{\zeta (5)}{2 \cdot 16}  l_s^4 (s^2 + t^2 + u^2) + \ldots \Big)
{\cal{R}}^4 ,\ee
thus leading to
\be \label{finalcont}
A_4^{\rm total}  =  \frac{\kappa_{11}^4 \hat{K}}{(2 \pi)^{11} l_{11}^3} \Big[ \frac{\pi^3}{32} \zeta (5)
e^{-2\phi^A} + \frac{\pi^4}{6!} T_2 E_2 (U,\bar{U})^{SL(2,\mathbb{Z})} 
+ \frac{\pi^7}{3 \cdot 6!} e^{2\phi^A}  + \ldots \Big] l_s^4 (s^2 +  t^2 + u^2).\ee
 
Thus, we see that \C{finalcont} leads to terms in the type IIB effective action given by 
\be \label{checktotact}
l_s^4 \int d^8 x \sqrt{-g_8} \Big[ (e^{-2\phi} V_2 ) 2 \zeta (5) + \frac{8 \zeta(4)}{\pi^3}
E_{2} (T,\bar{T})^{SL(2,\mathbb{Z})} U_2^2 + (e^{-2\phi} V_2 )^{-1} \frac{8 \zeta (4)}{3} U_2^2 \Big],\ee
where we have used $\zeta (4) = \pi^4/90$. To see that it reproduces some of the terms in \C{pertexp},
we keep the leading terms in $U_2$ in $F_0, F_1$, and $F_2$ in \C{checkval} giving us\footnote{We drop the
$E_2 (T,\bar{T})^{SL(2,\mathbb{Z})}$ term in $F_2$.}
\be F_0 = 2 \zeta (5) , \quad F_1 = \frac{8 \zeta(4)}{\pi^3}
E_{2} (T,\bar{T})^{SL(2,\mathbb{Z})} U_2^2 + \ldots, \quad F_2 = \frac{8 \zeta (4)}{3} U_2^2 + \ldots, \ee
which precisely matches \C{checktotact}. Thus the supergravity analysis provides some more
evidence for the proposed modular form.

\section{Decompactifying to nine dimensions}

We now decompactify the $D^4 {\cal{R}}^4$ interaction to nine dimensions to see what structure it gives, 
and also to make some further consistency checks. We define
\be T_2 = r_\infty r_B , \quad U_2 = \frac{r_\infty}{r_B} ,\ee 
where $r_\infty$ is the direction that is being decompactified. Here $r_\infty$ and $r_B$ are the radii
of $T^2$ in the string frame. Now let us take the limit $r_\infty \rightarrow \infty$, so that
$T_2, U_2 \rightarrow \infty$. From \C{modform}, using \C{expSL2} and \C{expEs}, 
we see that the non--vanishing
terms in nine dimensions in the string frame are given by
\bea \label{ninedim}
l_s^3 \int d^9 x \sqrt{- g_9} \Big[ \frac{r_B}{\sqrt{\tau_2}} 
E_{5/2} (\tau,\bar{\tau})^{SL(2,\mathbb{Z})} + \frac{4 \zeta (4)}{\pi^2 \tau_2^{3/2} r_B^3}
E_{3/2} (\tau,\bar{\tau})^{SL(2,\mathbb{Z})} \non \\
+ \frac{8}{\pi^2} \zeta (3) \zeta (4) r_B^3 + 
\frac{16}{\pi^3} \zeta (4)^2 r_\infty^3 \Big] D^4 {\cal{R}}^4, \eea
where we have set $ l_s \int d^8 x \sqrt{- g_8} r_\infty = \int d^9 x \sqrt{-g_9}$. 
Note that the last term in \C{ninedim} is divergent in the limit $r_\infty \rightarrow \infty$.
However the existence of this kind of term is crucial for the consistency of the theory. The full
effective action of type IIB string theory on $T^2$ contains terms analytic (like the 
$(s^2 + t^2 + u^2) {\cal{R}}^4$ term we have discussed) as well as 
non--analytic in the external momenta of the gravitons. In taking the decompactification limit
to go to nine dimensions, a part of the analytic terms diverges (\C{ninedim} contains only 
one such term among an infinite number of such diverging terms coming from the infinite
number of analytic terms). These diverging terms as well as the non--analytic terms must 
add up to give the massless square root threshold singularity in nine dimensions. A detailed analysis
of the corresponding divergence in ten dimensions obtained by taking $r_B \rightarrow \infty$
has been done in~\cite{Green:1999pu},
where it was shown that the diverging term involving $r_B^3$ in \C{ninedim}
adds up with other such terms, as well as the non--analytic terms to give the logarithmic threshold 
singularity in ten dimensions. So the term in \C{ninedim} that diverges as $r_\infty \rightarrow \infty$
is not a part of the $D^4 {\cal{R}}^4$ interaction in nine dimensions
(just like the $r_B^3$ term that diverges in the ten 
dimensional limit is not a part of the $D^4 {\cal{R}}^4$ interaction in ten dimensions
), and so we drop it from our analysis from now on.

Keeping only the terms that survive in the large $r_B$ limit, note that \C{ninedim}
reduces to
\be
l_s^3 \int d^9 x \sqrt{- g_9} r_B \Big[ e^{\phi_B/2} 
E_{5/2} (\tau,\bar{\tau})^{SL(2,\mathbb{Z})} + \frac{8}{\pi^2} \zeta (3) \zeta (4)  r_B^2
\Big] D^4 {\cal{R}}^4, \ee
which is precisely what has been obtained in~\cite{Green:1999pu}, which is a consistency check
of our modular form.
Also from \C{ninedim}, we see that the perturbative contribution to the scattering amplitude
is given by (upto an irrelevant numerical factor)
\be \label{pertnine}
\Big[ 2 \zeta (5) (r_B e^{- 2\phi_B}) + \frac{8}{\pi^2} \zeta (3) \zeta (4) \Big( r_B^3 
+ \frac{1}{r_B^3} \Big) + \frac{8}{3} \zeta (4) (r_B  
e^{- 2\phi_B})^{-1} \Big( r_B^2 
+ \frac{1}{r_B^2} \Big)\Big] (s^2 + t^2 + u^2) {\cal{R}}^4 , \ee
where the three terms give tree level, one loop and two loop contributions respectively. 
Now using the relations
\be r_B = r_A^{-1} , \quad e^{-\phi_B} = r_A e^{-\phi_A} , \ee
to go to type IIA string theory, from \C{pertnine} we see that the perturbative contributions 
are the same in either theory, which should be the case. 
As another consistency check, we now show that the tree level and 
one loop contributions in \C{pertnine} match the result using string perturbation theory. 

From \C{totalcont}, for $n=1$, we see that the amplitude in nine dimensions is given by 
\be \Big[ 2 \zeta (5) r_B e^{- 2\phi_B} + \hat{I}_{1} \Big] l_s^4 (s^2 + t^2 + u^2) {\cal{R}}^4, \ee
where the one loop contribution $\hat{I}_1$ is given by
\be \hat{I}_1 = \frac{4 r_B}{\pi} \sum_{m,n \in \mathbb{Z}} 
\int_{{\cal{F}}_L} \frac{d^2 \Omega}{\Omega_2^2} 
e^{-\pi r_B^2 \vert m + n \Omega\vert^2/\tau_2} E_{2}
(\Omega, \bar\Omega)^{SL(2,\mathbb{Z})}, \ee
where ${\cal{F}}_L$ is the restricted fundamental domain of $SL(2,\mathbb{Z})$ as before. This 
integral can be simplified leading to~\cite{McClain:1986id,O'Brien:1987pn,Ditsas:1988pm}
\bea \label{ninepert}
\hat{I}_1 = \frac{4 r_B}{\pi} \Big[ \int_{{\cal{F}}_L} \frac{d^2 \Omega}{\Omega_2^2}
+ \sum_{m \in \mathbb{Z} , m \neq 0} 
\int_{-1/2}^{1/2} d \Omega_1 \int_0^\infty \frac{d\Omega_2}{\Omega_2^2} e^{-\pi r_B^2 m^2 /\Omega_2}
\Big] E_{2} (\Omega, \bar\Omega)^{SL(2,\mathbb{Z})}. \eea

Now the first term in \C{ninepert} is 
proportional to $I_1$ in \C{I1cont}, and thus vanishes. 
Using \C{needdef} and doing the other integral, we get
\be \hat{I}_1 = \frac{8}{\pi^2} \zeta (3) \zeta (4) \Big( r_B^3 
+ \frac{1}{r_B^3} \Big) ,\ee
thus giving us \C{pertnine}. In the nine dimensional Einstein frame,
we see that \C{ninedim} equals
\be \label{Einsnine}
l_s^3 \int d^9 x \sqrt{- \hat{g}_9} \Big[ \xi^5 
E_{5/2} (\tau,\bar{\tau})^{SL(2,\mathbb{Z})} + \frac{4 \zeta (4) }{\pi^2} \xi^{-9}
E_{3/2} (\tau,\bar{\tau})^{SL(2,\mathbb{Z})} 
+ \frac{8}{\pi^2} \zeta (3) \zeta (4) \xi^{12} \Big] \hat{D}^4 {\hat{\cal{R}}}^4, \ee
where the hatted indices signify quantities in the Einstein frame, and $\xi^7 = r_B^2 \sqrt{\tau_2}$. Thus
from \C{Einsnine}, the $SL(2,\mathbb{Z}) \times \mathbb{R}^+$ U--duality symmetry of the 
$\hat{D}^4 {\hat{\cal{R}}}^4$ interaction is nine dimensions is manifest\footnote{Let 
us make some comments about the possible modular form for the $D^4 {\cal{R}}^4$ interaction
in lower dimensions. In dimensions lower than eight, the coset manifold ${\cal{M}} = G/H$ 
which parametrizes the scalars in the supergravity action is 
irreducible (see~\cite{Cremmer:1980gs,Julia:1980gr}, for example). 
Here $G$ is a 
non--compact group, and $H$ is its maximal compact subgroup. The conjectured U--duality group
is $\hat{G}$, the discrete version of $G$. 
Thus in the Einstein frame the relevant term in the supergravity action is given by 
\be 
S \sim \frac{1}{l_s^{8 -d}} \int d^{10-d} x \sqrt{-\hat{g}_{10-d}}  
{\rm Tr} (\p_\mu M \hat{\p}^\mu M^{-1}) ,\ee
where $M$ parametrizes ${\cal{M}}$. Based on the $D^4 {\cal{R}}^4$ interaction in ten
dimensions as well as the modular form we propose, it is conceivable that the 
U--duality invariant modular form in lower dimensions is given by
\be E_{5/2} (M)^{\hat{G}}  = \sum'_{m_i} 
\Big( m_i M_{ij} m_j \Big)^{-5/2}. \ee}.

It would be interesting to prove or disprove the modular form for the $D^4 {\cal{R}}^4$ 
interaction we have proposed, as well as to construct U--duality invariant modular forms for the four
graviton amplitude in toroidal compactifications of type IIB string theory to lower dimensions. 
It might be possible to construct the modular form for the $D^6 {\cal{R}}^4$ 
interaction in eight dimensions along the lines in this paper, although the analysis will
get more complicated because the ten dimensional modular form satisfies a Poisson equation on the 
fundamental domain of $SL(2,\mathbb{Z})$~\cite{Green:2005ba}. In trying to construct these modular forms,
it might be useful to consider eleven dimensional supergravity compactified on torii, and consider
four graviton scattering in this background. Though this will not account for the various non--perturbative
contributions like membrane instantons for M theory on $T^3$, 
it might give hints about the various U--duality invariant modular forms. 
In general, understanding the role of modular forms in toroidal compactifications
of M theory that preserve all the thirty two supersymmetries is useful. At least some aspects of
constructing them might not depend on a precise definition of the microscopic degrees of freedom
of M theory, and might be completely determined based on the constraints of supersymmetry and 
U--duality invariance. Thus constructing them might shed some light on the fundamental degrees of
freedom of M theory.   
  
\section*{Acknowledgements}

I would like to thank J.~Maldacena for useful comments. 
The work of A.~B. is supported by NSF Grant No.~PHY-0503584 and the William D. Loughlin membership.

\section{Appendix}

In the two appendices below, we write down explicit expressions for the Eisenstein series of 
$SL(2,\mathbb{Z})$ and $SL(3,\mathbb{Z})$ that are useful in the main text.
 
\appendix

\section{The Eisenstein series for $SL(2,\mathbb{Z})$}

The Eisenstein series of order $s$ for $SL(2,\mathbb{Z})$ 
is defined by
\bea \label{expSL2}
E_{s} (T,\bar{T})^{SL(2,\mathbb{Z})}  &=& \sum_{(p,q) \neq (0,0)} 
\frac{T_2^{s}}{\vert p + q T\vert^{2s }} \non \\
&=& 2 \zeta(2s) T_2^s + 2 \sqrt{\pi} T_2^{1-s} \frac{\Gamma (s -1/2)}{\Gamma (s)}
\zeta (2s -1) \non \\ &&+ \frac{2 \pi^s \sqrt{T_2}}{\Gamma (s)} \sum_{m_1 \neq 0 ,m_2 \neq 0} 
\Big\vert \frac{m_1}{m_2} \Big\vert^{s - 1/2} K_{s - 1/2} (2\pi T_2 \vert m_1 m_2
\vert) e^{2\pi i m_1 m_2 T_1}. \eea

Using the relations

\be \label{imprel}
\zeta (2s -1 ) \Gamma (s - \frac{1}{2}) = \pi^{2s - 3/2} \zeta (2 - 2s) \Gamma (1-s) , \ee
and 
\be K_s (x) = K_{-s} (x) , \ee
we see that
\be \label{relSL2}
\Gamma (s) E_{s} (T,\bar{T})^{SL(2,\mathbb{Z})}  = \pi^{2s - 1} \Gamma (1-s) 
E_{1-s} (T,\bar{T})^{SL(2,\mathbb{Z})} . \ee

\section{The Eisenstein series for $SL(3,\mathbb{Z})$}

The Eisenstein series of order $s$ for $SL(3,\mathbb{Z})$ in the fundamental representation
is defined by
\bea  \label{expSL3}
E_s (M)^{SL(3,\mathbb{Z})}  &=& \sum'_{m_i} \Big( m_i M_{ij} m_j \Big)^{-s} \non \\
& =&\sum'_{m_i} \nu^{-s/3} \Big( \frac{\vert m_1 + m_2 \tau + m_3 B
\vert^2}{\tau_2}+\frac{m_3^2}{\nu} \Big)^{-s}, \eea
where $m_i$ are integers, and the sum excludes $\{ m_1, m_2, m_3 \} = \{ 0,0,0 \}$.
The integers $m_i$ transform in the anti--fundamental representation of $SL(3,\mathbb{Z})$. The
matrix $M_{ij}$ has entries given by \C{matrix}.

Using the integral representation
\bea  \label{expE}
E_s (M)^{SL(3,\mathbb{Z})}  =
\frac{\nu^{-s/3} \pi^s}{\Gamma (s)} \int_0^\infty \frac{d t}{t^{s + 1}} \sum'_{m_i} 
e^{-\pi ( \vert m_1 + m_2 \tau + m_3 B
\vert^2/\tau_2+ m_3^2/\nu )/t }, \eea
we can evaluate \C{expE} to get that
\bea \label{expEs}
&&E_s (M)^{SL(3,\mathbb{Z})} = 2 ( \tau_2^2 V_2)^{2s/3} \zeta (2s) + \frac{\sqrt{\pi} \Gamma (s -  
1/2)}{\Gamma (s)}
(\tau_2^2 V_2)^{1/2 - s/3} E_{s -1/2} (T,\bar{T})^{SL(2,\mathbb{Z})} \non \\&&
+ \frac{2 \pi^s}{\Gamma(s)} \tau_2^{s/3 + 1/2} V_2^{2s/3} 
\sum_{m_1 \neq 0 ,m_2 \neq 0} 
\Big\vert \frac{m_1}{m_2} \Big\vert^{s - 1/2} K_{s - 1/2} (2\pi \tau_2 \vert m_1 m_2
\vert) e^{2\pi i m_1 m_2 \tau_1} \non \\  &&+ \frac{2 \pi^s}{\Gamma(s)} \tau_2^{1 -2s/3} V_2^{1 - s/3} 
\sum_{m_1 \neq 0, m_3 \neq 0, m_2} \Big\vert \frac{m_2 - m_1 \tau}{m_3} 
\Big\vert^{s-1} K_{s-1} (2\pi \vert m_3
( m_2 - m_1 \tau) \vert V_2) \non \\ && ~~~~~~~~~~~~~~~~~~~~~~~~~~~~~~~~
~~~~~~~~~~~~~~~~~~~~~~~~~~~~~~~~~~~~~\times e^{2\pi i m_3 (m_1 B_R + m_2 B_N)}. \eea

We can also define the Eisenstein series of order $s$ in the anti--fundamental representation
by
\bea  \label{expEanti}
E_s (M^{-1})^{SL(3,\mathbb{Z})}  = \sum'_{\hat{m}_i} \Big( \hat{m}_i M^{ij} \hat{m}_j \Big)^{-s}, \eea
where $\hat{m}_i$ transforms in the fundamental representation of $SL(3,\mathbb{Z})$. Now using the 
result 
\be \label{needresum}
\sum'_{\hat{l}_i} e^{-\pi \s G^{ij} \hat{l}_i \hat{l}_j} = \s^{-3/2} \sqrt{{\rm det} G} \sum'_{l_i} 
e^{-\pi G_{ij} l_i l_j/\s} \ee
for invertible matrices, which can be derived using Poisson resummation, we get that
\be \label{relfundantifund}
E_s (M^{-1})^{SL(3,\mathbb{Z})}  = E_{3/2 -s} (M)^{SL(3,\mathbb{Z})}  . \ee

Thus there is a simple relationship between the Eisenstein series for the fundamental and the
anti--fundamental representations.

%\newpage  
%\bibliographystyle{amsunsrt-es}
%\bibliography{myrefs}
%\bibliographystyle{utphys}
%\bibliography{myrefs}

%\end{thebibliography}
%\end{document}

%\begin{thebibliography}{10}

%\ifx\undefined\bysame
%\newcommand{\bysame}{\leavevmode\hbox to3em{\hrulefill}\,}
%\fi

%\providecommand{\href}[2]{#2}\begingroup\raggedright\begin{thebibliography}{10}

%\end{thebibliography}\endgroup
%\end{thebibliography}
\providecommand{\href}[2]{#2}\begingroup\raggedright\endgroup

\end{document}